\documentclass[ aps,%
floatfix,%
final,%
notitlepage,%
oneside,%
twocolumn,%
nobibnotes,%
nofootinbib,%
superscriptaddress,%
showpacs,%
]%
{revtex4}

\usepackage{graphicx}
\usepackage{amssymb}
\newcommand{\eqref}[1]{(\ref{#1})}

\begin{document}

\title{Double vector quarkonia production in exclusive Higgs boson decays}

\author{Kartvelishvili, V.}\email{V.Kartvelishvili@lancaster.ac.uk}
\affiliation{Lancaster University, Lancaster, UK}

\author{Luchinsky A.V.}\email{Alexey.Luchinsky@ihep.ru}
\affiliation{Institute for High Energy Physics, Protvino, Russia}

\author{Novoselov A.A.}\email{Alexey.Novoselov@cern.ch}
\affiliation{Institute for High Energy Physics, Protvino, Russia}

\begin{abstract}
Partial decay widths and branching fractions are calculated for the exclusive decays of the Standard Model Higgs boson into a pair of vector quarkonium states, $ H\to J/\psi J/\psi,  H\to \Upsilon \Upsilon, H\to J/\psi \phi, H\to J/\psi \Upsilon$, with relativistic corrections due to quark motion in mesons taken into account.
\end{abstract}


\maketitle

% main text
\section{Introduction}

The necessary ingredient of the Standard Model (SM) is the Higgs boson --- a scalar particle whose interactions are believed to generate masses of intermediate vector bosons and fundamental fermions. However, numerous attempts to discover this particle experimentally have so far only resulted in the limits on its possible mass $M_H$, $114.4\,\mathrm{GeV}<M_{H}\lesssim182\,\mathrm{GeV}$ \cite{Amsler:2008zz}.

Decay modes that can be used to detect this particle depend on $M_{H}$. The coupling constant of the Higgs boson to another particle is known to be proportional to the mass of the latter, hence it is advantageous to search for Higgs boson decays into the heaviest of kinematically allowed particles. For instance, if $M_{H}$ is above $WW, ZZ$ production thresholds, $H\to WW,ZZ$ decays (so called ``gold-plated'' modes) can be used. Observation of such reactions is one of the main goals of the LHC physics programme.

However, the possibility of $M_{H}<2M_W$ cannot be excluded. In this case $b\bar b$ and $\tau^+\tau^-$ modes are dominant, but large backgrounds makes these decays harder to observe. In search for an exclusive Higgs decay mode with a good signature, in
\cite{Keung:1983ac,Doroshenko:1987nj,Kartvelishvili:1988pu}
decays of the Higgs boson into a pair of heavy quarkonia ($H\to J/\psi J/\psi$, $\Upsilon\Upsilon$, $J/\psi\Upsilon$ etc.) were considered. The widths of such decays were found to be small, but their good signatures and high attainable mass resolutions could make these decays useful in certain circumstances.

The calculations in \cite{Keung:1983ac,Doroshenko:1987nj,Kartvelishvili:1988pu} were performed in the so-called $\delta$-approximation, i.e. they did not take into account relativistic corrections caused by the internal motion of quarks in the vector mesons. Since then, however, several theoretical and experimental studies of double charmonium production in exclusive processes have shown that the relativistic corrections can significantly alter the rates of these processes
%In the works \cite{Abe:2002rb,Bondar:2004sv,Braguta:2005gw,Braguta:2005kr,Braguta:2006nf} it was shown, that the cross sections of double charmonia production in electron-positron annihilation are increased by about an order of magnitude.
(see \cite{Abe:2002rb,Bondar:2004sv,Braguta:2005gw,Braguta:2005kr,Braguta:2006nf} for more details).
In particular, the cross section of the reaction $e^+e^-\to J/\psi\eta_c$ is increased by about an order of magnitude \cite{Abe:2002rb,Bondar:2004sv,Braguta:2005gw}. Another example is the decay $\chi_{b}\to J/\psi J/\psi$: in \cite{Braguta:2005gw} it was shown that by taking the internal quark motion into account, the width of this decay increases by a factor of 3.

In this paper we study the influence of relativistic corrections on exclusive decays of the SM Higgs boson into a pair of vector quarkonia.
The rest of the paper is organized as follows: in the next section we briefly describe the formalism used, and present the distribution amplitudes for various vector quarkonium states. In section \ref{sec:HVV} analytical expressions are given for $H\to V_{1}V_{2}$ decay amplitudes corresponding to leading contributing subprocesses. In section \ref{sec:NUM} we present numerical results for the Higgs decays into $V_{1}V_{2}=J/\psi J/\psi$, $\Upsilon(1S)\Upsilon(1S)$, $J/\psi \phi$ and $J/\psi\Upsilon(1S)$ final states, and compare them to the results obtained within the $\delta$-approximation. Our conclusions are given in the final section.

\section{Distribution amplitudes\label{sec:DA}}

Consider double quarkonia production in exclusive Higgs boson decays
\begin{eqnarray*}
H & \to & V_{1}\left(p_{1},\lambda_{1}\right)V_{2}\left(p_{2},\lambda_{2}\right)
\end{eqnarray*}
where $p_{1,2}$ and $\lambda_{1,2}$ are momenta and helicities of the two vector mesons, respectively.

In what follows, we restrict ourselves to the leading twist contribution.
It can be shown that the final vector mesons in these decays are mainly longitudinally polarized, i.e
$\lambda_{1,2}=0$. Indeed, consider the explicit form of the polarization vector of a vector meson with energy $E$ and momentum $p$. For a longitudinally polarized vector meson one has
\begin{eqnarray}
  \epsilon_\mu(\lambda=0) &=& \left\{\frac{p}{M_V} , 0 , 0 , \frac{E}{M_V} \right\} \sim \frac{M_H}{M_V},
  \label{eq:epsL}
\end{eqnarray}
where $M_V$ and $M_H$ are masses of the vector meson and the Higgs boson, respectively. On the other hand, for a transversely polarized vector meson,
\begin{eqnarray*}
\epsilon(\lambda=\pm 1) &=& \left\{0,\frac{1}{\sqrt 2} , \pm\frac{i}{\sqrt 2} , 0\right\} \sim 1.
\end{eqnarray*}
This vector is suppressed by a small factor $\sim {\cal{O}}\left( M_V/M_H \right)$ in comparison with the expression \eqref{eq:epsL}, so in the following we will only consider the case $\lambda_1=\lambda_2=0$.

The transition of a quark-antiquark pair into a longitudinally polarized vector meson is described by the expression
\begin{eqnarray}
\left\langle
V(p,\lambda=0)\left|\bar{q}_{\alpha}^{i}(z) q_{\beta}^{j}(-z)\right|0\right\rangle
& = &
\frac{f}{4}\frac{\delta^{ij}}{3}\left(\hat{p}\right)_{\alpha\beta}
\times\nonumber\\ &\times&
\intop_{0}^{1}dx\varphi(x) e^{i(2x-1)pz},\label{eq:proj}
\end{eqnarray}
where $\alpha$ ($\beta$) and $i$ ($j$) are spinor and color indices of the quark (antiquark), respectively, while $x$ is the quark momentum fraction with respect to the meson momentum. The constant $f$ can be determined from the leptonic width of the vector meson:
\begin{eqnarray}
\Gamma\left(V\to e^{+}e^{-}\right) & = & \frac{4\pi\alpha^{2}}{3}e_{q}^{2}\frac{f^{2}}{M},\label{eq:f}
\end{eqnarray}
where $\alpha$ is the fine structure constant, $e_q$ is the quark charge ($2/3$ for $J/\psi$, $-1/3$ for $\phi$ and $\Upsilon$), and $M$ is the mass of quarkonium.

A distribution amplitude $\varphi(x)$, describing the internal motion of the heavy quark-antiquark pair inside  their bound state, can generally be written as a series of Gegenbauer polynomials $C_n^{3/2}$ \cite{Chernyak:1983ej}:
\begin{eqnarray}
  \varphi(x,\mu) &=& 6x\bar x\left[
    1+\sum\limits_{n=2,4,\dots} a_n(\mu) C_n^{3/2} (x-\bar x)
  \right],
  \label{eq:phi}
\end{eqnarray}
where we have introduced a simplifying notation $\bar x=1-x$. The dependence of this distribution on energy scale $\mu$ is described by the QCD evolution of the moments $a_n(\mu)$:
\begin{eqnarray}
  a_n(\mu) &=& \left(\frac{\alpha_s(\mu)}{\alpha_s(\mu_0)}\right)^{\gamma_n} a_n(\mu_0),
  \label{eq:a}
\end{eqnarray}
with the anomalous dimensions $\gamma_n$ defined by
\begin{eqnarray}
  \gamma_n &=& \frac{4}{3b_0}\left(1-\frac{2}{(n+1)(n+2)}+4\sum\limits_{j=2}^{n+1}\frac{1}{j} \right),
  \label{eq:gammaN}\\
  b_0 &=& 11-\frac{2}{3}n_f, %\nonumber
\end{eqnarray}
where $n_f$ is the number of active quark flavours. These anomalous dimensions are positive, so in the limit $\mu\to\infty$ (or, equivalently, for light mesons) moments (\ref{eq:a}) tend to zero. As a result, in this limit distribution amplitude (\ref{eq:phi}) tends to its asymptotic
form $\varphi_{a}(x)=6x\bar x$. We use this form of distribution amplitude for the $\phi$ meson.

However, for heavier quarkonia $J/\psi$ and $\Upsilon$ and for the scales $\mu=M_H\sim 100\div 250$ GeV, the asymptotic limit is not reached yet. Following \cite{Braguta:2007fh}, for $J/\psi$ we use the  distribution amplitude
\begin{eqnarray}
  \varphi_{J/\psi}(x,m_c) &=& c(\beta)\left( 1-(x-\bar x)^2\right)\times\nonumber\\
    &\times& \exp\left\{ -\frac{\beta}{1-(x-\bar x)^2} \right\}
    \label{eq:phiExp}
\end{eqnarray}
with $\beta\approx 3.8$ and the factor $c(\beta)$ fixed from the normalization condition
\begin{eqnarray*}
  \int\limits_0^1\varphi(x)dx &=& 1
\end{eqnarray*}
Note that this function is fairly close to a simpler expression \cite{Kartvelishvili:1985ac,Gershtein:2006ng}
\begin{eqnarray}
  \varphi_{J/\psi}(x) &\sim& x^{-\alpha_\psi}{\bar x}^{-\alpha_\psi}
  \label{eq:psiR}
\end{eqnarray}
where $\alpha_\psi \simeq -3$ is the intercept of the Regge trajectory corresponding to charmonium.
For the wave function of $\Upsilon$ we use a parametrization similar to \eqref{eq:psiR} with the intercept $\alpha_\Upsilon=-9$ \cite{Gershtein:2006ng}. The typical scale $\mu$ for these distribution amplitudes is the corresponding quark mass. Using formulae (\ref{eq:phi}), (\ref{eq:a}) and (\ref{eq:gammaN}), their evolution  to $\mu=M_H$ can be easily calculated.

\section{$H\to V_{1}V_{2}$\label{sec:HVV}}

\begin{figure}
\begin{centering}
\includegraphics[width=7cm]{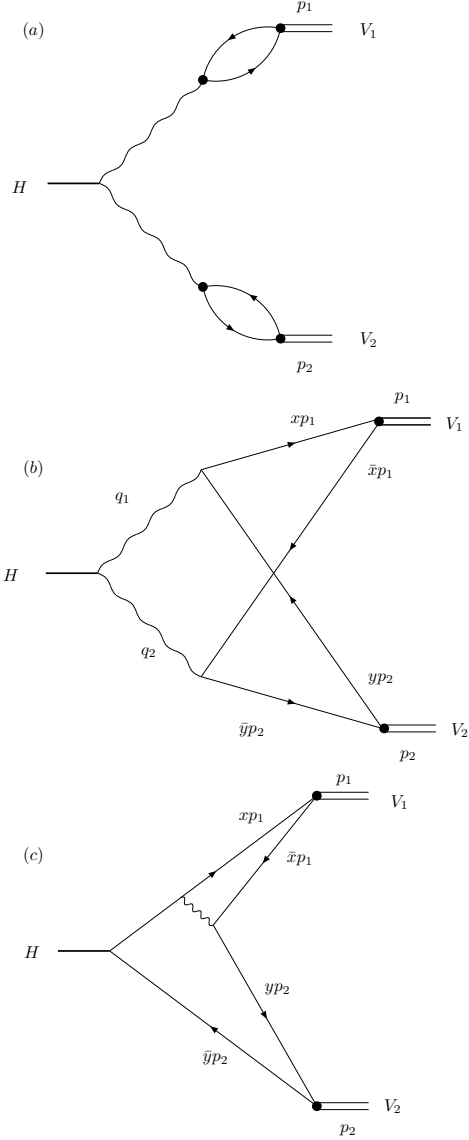}
\par\end{centering}
\caption{Typical diagrams describing $H\to V_{1}V_{2}$ decays.
\label{diags}}
\end{figure}

Typical diagrams that give leading contributions to $H\to V_{1}V_{2}$ decays are shown in fig. \ref{diags}.

%Let us first consider the diagram shown in fig. \ref{diags}a. It can be easily shown, that the corresponding amplitude is suppressed by vector meson mass squared. For this reason we neglect such diagrams.

For the diagram shown in fig.~\ref{diags}a, only $Z$ will contribute as the intermediate virtual bosons, as the final mesons should be neutral and colourless. On the other hand, in this subprocess the two final mesons can be composed from different type of quarks. The corresponding amplitude is
\begin{eqnarray}
\mathcal{M}^{(ZZr)} & = & -\left[\frac{e}{\sin2\vartheta_{W}}\right]^{3}\frac{a_{1}f_{1}}{D_{Z}(M_{1}^{2})}\frac{a_{2}f_{2}}{D_{Z}(M_{2}^{2})}M_{H}^{2}M_{Z}.
\label{eq:ampZZr}
\end{eqnarray}
Here $\vartheta_{W}$ is the Weinberg angle,
$D_{Z}(M)=M^{2}-M_{Z}^{2}+iM\Gamma_{Z}$ is the inverse propagator of the virtual $Z$-boson,
and constants $a_{1,2}$ are defined as
\begin{eqnarray*}
a_{i} & = & \pm\frac{1}{2}-2e_{q}\sin^{2}\vartheta_{W}
\end{eqnarray*}
with $+$ ($-$) sign used for $c$ ($s,b$) quarks.
%and $-$ Plus and minus signs in this expression correspond to mesons that are built from up-type ($u$, $c$, $t$) and down-type ($d$, $s$,$b$) quarks respectively.
One can see that this amplitude is expressed through experimentally observable quantities only, and hence does not depend on distribution functions $\varphi(x)$. Note also that this amplitude increases with increasing Higgs boson mass.

The situation is rather different for the subprocess described by the diagram in fig.~\ref{diags}b. In this case, either $Z$, $W$ or gluons can be used as virtual particles. Their momenta can be calculated as
\begin{eqnarray*}
q_{1} & = & xp_{1}+yp_{2},\qquad q_{2}=\overline{x}p_{1}+\overline{y}p_{2},
\end{eqnarray*}
where $p_{1,2}$ are the momenta of final vector mesons, $x$ and $y$ are momentum fractions of quarks inside $V_{1}$ and $V_{2}$, respectively, while $\overline{x}\equiv 1-x$ and $\overline{y} \equiv 1-y$. So, neglecting the masses of the quarkonia compared to $M_{W,Z}$ and $M_H$, the virtualities of the intermediate bosons are
\begin{eqnarray*}
q_{1}^{2} & = & 2xy(p_{1}p_{2})=xyM_{H}^{2},\qquad q_{2}^{2}=\overline{x}\,\overline{y}M_{H}^{2}.
\end{eqnarray*}
Clearly, in contrast with the previous diagram, these amplitudes depend on the distribution functions $\varphi_{1,2}(x)$. In addition, they do not increase with the increasing Higgs boson mass.

In the case of $ZZ$ decay, the quarks in mesons should be the same. The amplitude is equal to
\begin{eqnarray}
\mathcal{M}^{(ZZ)} & = & -\frac{1}{3}\left[\frac{e}{\sin2\vartheta_{W}}\right]^{3}b_{q}M_{H}^{2}f_{1}f_{2}M_{Z}
    \times\nonumber\\ &\times&
    \intop_{0}^{1}dx\, dy\,\frac{\varphi_{1}(x)}{D_{Z}(xyM_{H}^{2})}\frac{\varphi_{2}(y)}{D_{Z}(\overline{x}\,\overline{y}M_{H}^{2})},
\label{eq:ampZZ}
\end{eqnarray}
where
\begin{eqnarray*}
b_{q} & = & \frac{1}{2}\mp2e_{q}\sin^{2}\vartheta_{W}+4e_{q}^{2}\sin^{4}\vartheta_{W}
\end{eqnarray*}
and the upper (lower) sign corresponds to mesons built from up- (down-) type quarks. In the limit of large $M_H$ this amplitude remains constant. In the region $M_{H}\approx2M_{Z}$, however, there is a noticeable peak, because in this case, due to the specific kinematics of the quarkonium formation, both intermediate $Z$-bosons are bound to be close to the mass shell.

If the decay in fig.~\ref{diags}b is mediated by a pair of $W$ bosons,
quarks in different mesons should be different, with one of the mesons built from up-type quarks, while the other --- from down-type quarks. Here we consider $J/\psi \phi$  and $\Upsilon J/\psi$ final states. The amplitude of such a decay is equal to
\begin{eqnarray}
\mathcal{M}^{(WW)} & = & -\frac{1}{24}\frac{e^{3}\cot\vartheta_{W}}{\sin^{2}\vartheta_{W}}\left|V_{12}\right|^{2}M_{H}^{2}f_{1}f_{2}M_{Z}
    \times\nonumber\\ &\times&
\intop_{0}^{1}dx\, dy\frac{\varphi_{1}(x)}{D_{W}(xyM_{H}^{2})}\frac{\varphi_{2}(y)}{D_{W}(\overline{x}\,\overline{y}M_{H}^{2})},
\label{eq:ampWW}
\end{eqnarray}
where $V_{12}$ is the respective CKM matrix element, while $D_{W}(M)=M^{2}-M_{W}^{2}+iM\Gamma_{W}$ is the inverse propagator of a virtual $W$-boson. This amplitude also has a peak, this time at $M_{H}\approx2M_{W}$.

In the case of $gg$ intermediate state, quark content of the vector mesons should again be the same. In the Standard Model there is no $H\to gg$ coupling at tree level, but it appears at higher orders
when loop diagrams are taken into account. The main contribution to this effective vertex comes for the top quark loop, for which we use the expression (see, e.g., \cite{Okun:1982ap}) %\cite{hgg}
\begin{eqnarray*}
%\mathcal{M}(H\to gg) & = & N_{H}\frac{\alpha_{s}(M_{H})}{6\pi}\frac{e}{\sin2\theta_W %M_{Z}}G_{\mu\nu}^{1a}G_{\mu\nu}^{2a},
\mathcal{M}(H\to gg) & = & \frac{\alpha_{s}(M_{H})}{6\pi}\frac{e}{\sin2\theta_W M_{Z}}G_{\mu\nu}^{1a}G_{\mu\nu}^{2a},
\end{eqnarray*}
where
%$N_{H}$ is number of heavy quarks, i.e. quarks with mass greater than $M_{H}/2$ (for the Higgs mass region concerned only $t$-quark agrees with this condition, so $N_{H}=1$), and
the tensor $G_{\mu\nu}$ is defined according to
\begin{eqnarray*}
G_{\mu\nu}^{ia} & = & q_{i\mu}\epsilon_{i\nu}-q_{i\nu}\epsilon_{i\mu}.
\end{eqnarray*}
Strictly speaking, the value of the strong coupling constant in this vertex depends on gluon virtualities, but this dependence is only logarithmic and we do not take it into account. Using this effective vertex, we obtain the following expression for $H\to gg\to V_{1}V_{2}$ decay amplitude:
\begin{eqnarray}
\mathcal{M}^{(gg)} & = &
    -\frac{\sqrt{2}}{27}\frac{e\tan\vartheta_{W}\alpha_{s}(M_{H})}{M_{Z}}f_{1}f_{2}
    \times\nonumber\\ &\times&
    \intop_{0}^{1}dxdy\varphi_{1}(x)\varphi_{2}(y)\left[\frac{1}{x\overline{y}}+\frac{1}{y\overline{x}}\right].
\label{eq:ampGG}
\end{eqnarray}
Clearly, this amplitude only depends on the Higgs boson mass through $\alpha_s$.

There are other subprocesses contributing to the same double quarkonium final states, notably those
containing the tree-level vertex $H\to\bar q q$ shown in fig.~\ref{diags}c,
first calculated in \cite{Keung:1983ac}.
In such subprocesses, the two quarkonia must again be the same, and the amplitude can be written as
\begin{eqnarray*}
\mathcal{M}^{(qq)} &=& \frac{\pi\alpha_s}{9}\frac{M_V^2}{M_Z}\frac{f_1 f_2}{M_H^2}\int\limits_0^1 dx dy\frac{\varphi_1(x)\varphi_2(y)}{\bar x y}
  \left[ \frac{1}{\bar x} + \frac{1}{y} \right].
\end{eqnarray*}
Compared to other amplitudes considered above, in the Higgs mass range of interest
this amplitude is suppressed by a small factor $(M_V/M_H)^2$,
so we will neglect it in the following.

\section{Numerical results\label{sec:NUM}}

In order to obtain numerical results from eqs.
(\ref{eq:ampZZr}---\ref{eq:ampGG}),
%(\ref{eq:ampZZr}), (\ref{eq:ampZZ}), (\ref{eq:ampWW}) and (\ref{eq:ampGG}),
one needs the values for constants $f$ and the distribution amplitudes $\varphi(x)$.
The constants $f$, as determined from the leptonic decay widths $V\to e^{+}e^{-}$
using eq. (\ref{eq:f}), yield $f=200$ MeV, 400 MeV and 700 MeV for $\phi$, $J/\psi$ and $\Upsilon(1S)$ mesons, respectively. Our choices for the distribution amplitudes of these three vector mesons were given by eqs. (\ref{eq:phi}, \ref{eq:phiExp}, \ref{eq:psiR}), respectively.  After fixing these, calculations of partial decay widths for the decays $H\to V_{1}V_{2}$ are fairly straightforward.

\begin{figure}[t]
\begin{centering}
\includegraphics[width=8.8cm]{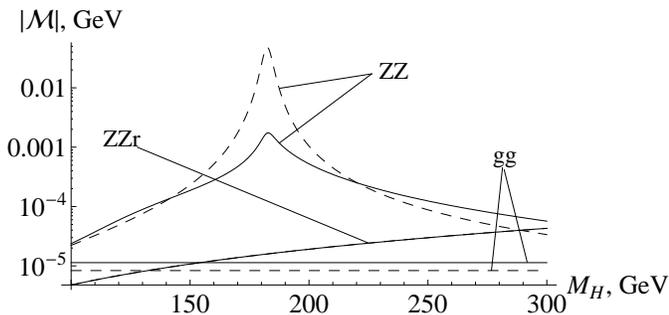}
\end{centering}
\caption{
Amplitudes of different subprocesses contributing to the decay $H\to J/\psi J/\psi$ versus the Higgs boson mass. Solid (dashed) lines stand for internal quark motion taken into account (neglected).
See text for further explanation.\label{fig:amp}}
\end{figure}

As an example, in fig. \ref{fig:amp} we show $M_H$-dependence of  various amplitudes contributing to the decay $H\to J/\psi J/\psi$.  The amplitudes calculated
with internal quark motion taken into account are shown with solid lines, while for dashed lines this motion is neglected. In $\delta$-approximation, the amplitude of the subprocess
$H\to ZZ\to J/\psi J/\psi$ has a prominent peak at $M_{H}=2M_{Z}$ (dashed line labelled $ZZ$). The reason for this peak, as explained above, is that at this value of the Higgs mass both virtual $Z$-bosons in fig. \ref{diags}b are on mass shell. When the internal quark motion is taken into account,
the quark momentum fractions $x$, $y$ are allowed to deviate from $1/2$, and the $Z$-bosons are off mass shell. As a result, the peak is washed out, while away from the peak the amplitude becomes somewhat larger (solid line). The amplitude of the
subprocess shown in fig. \ref{diags}a (solid line labelled $ZZr$) remains unchanged when the internal
quark motion is taken into account. As for the gluon-gluon subprocess (solid and dashed lines labelled $gg$), internal quark motion increases the corresponding amplitude roughly by a factor of 2.

\begin{figure}
\begin{centering}
\includegraphics[height=21cm,width=9cm]{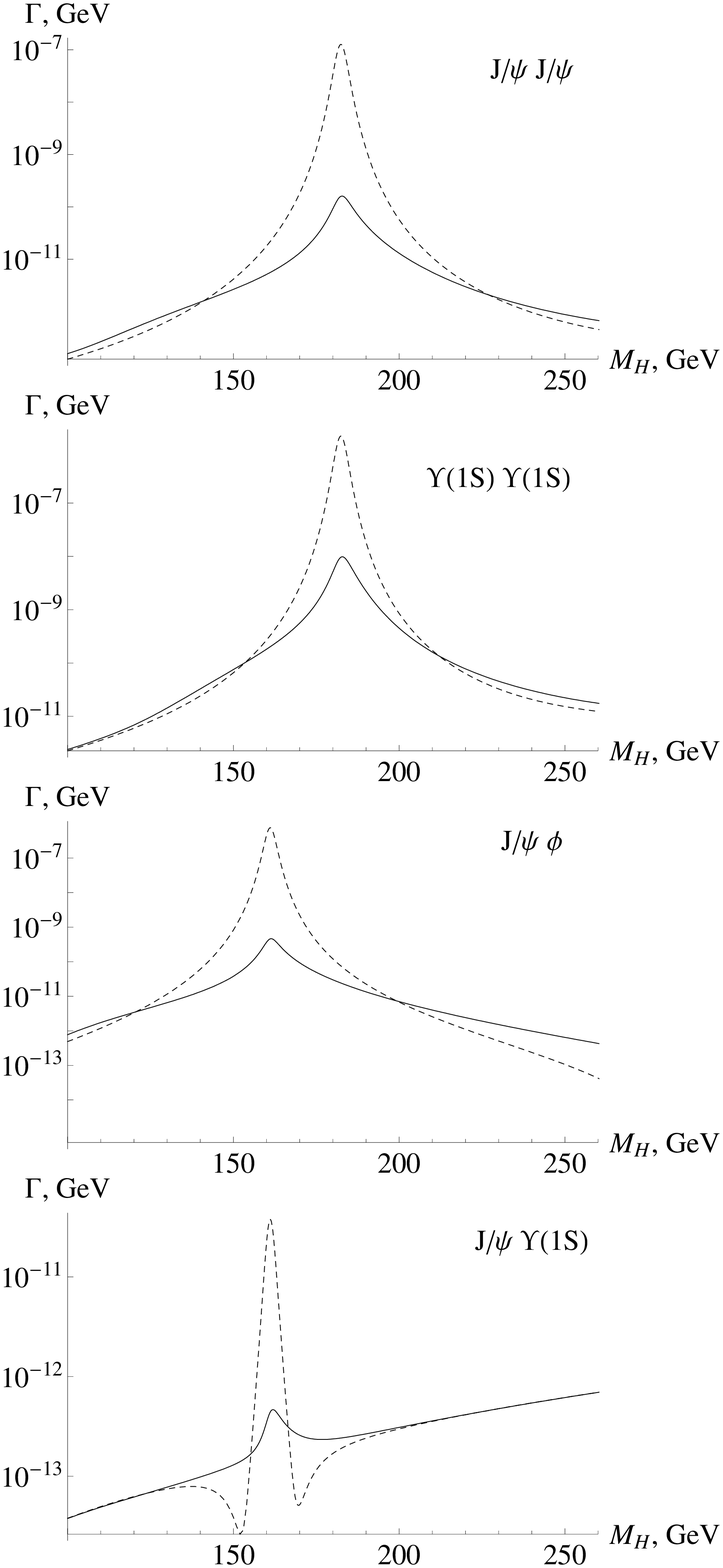}
\par\end{centering}
\caption{Partial decay widths for the exclusive decays $H\to V_{1}V_{2}$ versus the Higgs boson mass (in GeV) with internal quark motion taken
into account (solid lines) and neglected (dashed lines): from top to bottom,
the final states are
$J/\psi J/\psi$, $\Upsilon\Upsilon$, $J/\psi \phi$ and  $J/\psi\Upsilon$.
\label{fig:Gamma}}
\end{figure}

The main features of the above analysis also hold for other decays considered here.
In fig \ref{fig:Gamma} we present, from top to bottom,
the $M_H$-dependence of the partial widths of the decays $H\to J/\psi J/\psi$, $H\to\Upsilon\Upsilon$, $H\to J/\psi \phi$  and  $H\to J/\psi\Upsilon$, with the internal quark motion taken into account (solid lines) and neglected (dashed lines). The resonant peaks at $M_H = 2M_{Z,W}$, prominent in the $\delta$-approximation in all these cases, are significantly  smeared out once internal quark motion is accounted for. For Higgs boson masses away from these peaks, relativistic corrections increase the widths of the decays.

\begin{figure}
\begin{centering}
\includegraphics[height=21cm,width=9cm]{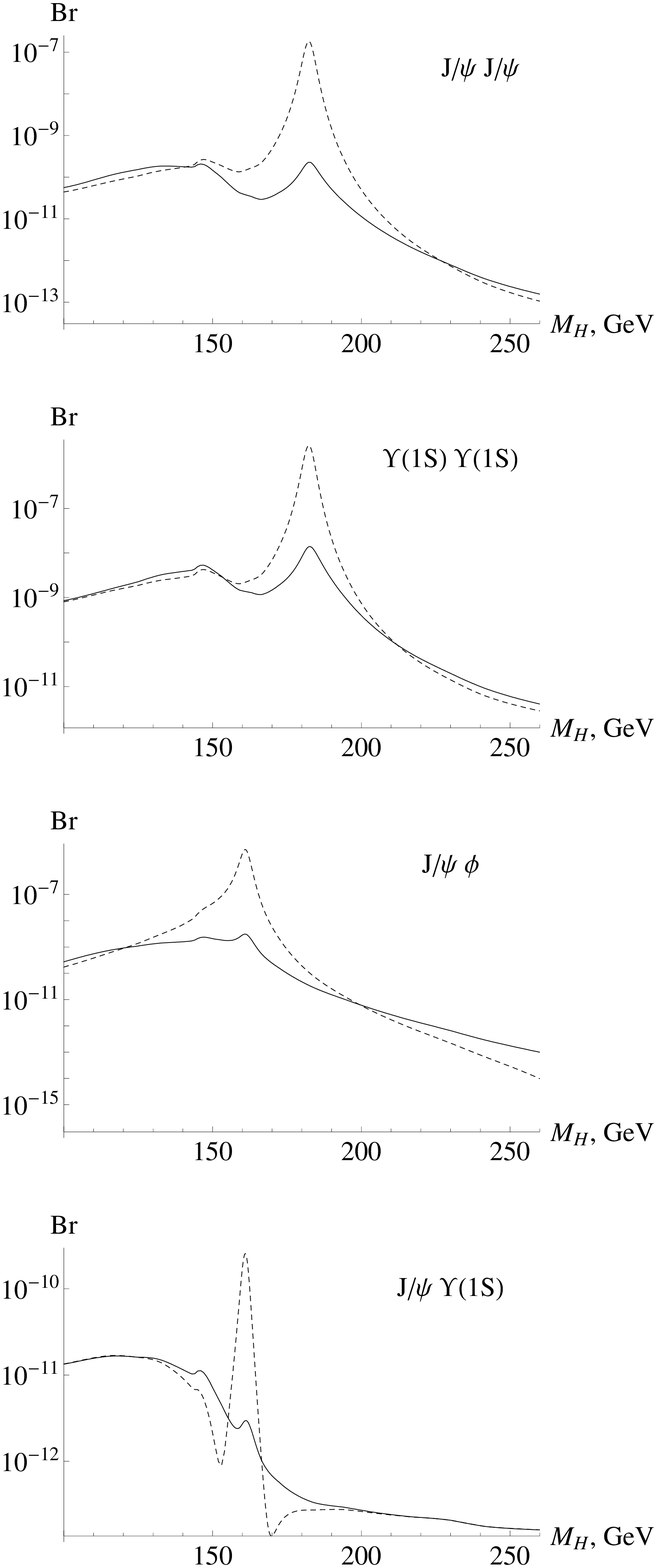}
\par\end{centering}
\caption{Branching fractions of the exclusive decays $H\to V_{1}V_{2}$ versus the Higgs boson mass,
with internal quark motion taken
into account (solid lines) and neglected (dashed lines): from top to bottom, the final states are
$J/\psi J/\psi$, $\Upsilon\Upsilon$, $J/\psi \phi$ and  $J/\psi\Upsilon$.
\label{fig:Br}}
\end{figure}

Using the total width of the Higgs boson calculated following \cite{Djouadi:1997yw}, we have calculated the branching fractions for the four decays under consideration, from top to bottom, $H\to J/\psi J/\psi$, $H\to\Upsilon\Upsilon$, $H\to J/\psi \phi$ and  $H\to J/\psi\Upsilon$. They are presented in figs. \ref{fig:Br}. It is clear that the branching fractions vary strongly depending on the Higgs boson mass, and that the internal quark motion changes these fractions significantly.

\section{Conclusion}
We have calculated
the partial decay widths and branching fractions for the exclusive decays of the Standard Model Higgs boson into a pair of vector quarkonium states, $H\to J/\psi \phi, H\to J/\psi, J/\psi, H\to J/\psi \Upsilon, H\to \Upsilon \Upsilon$, with relativistic corrections due to quark motion in mesons taken into account, and compared our results with previous calculations performed in $\delta$-approximation. We have found that
the characteristic increases of the respective decay probabilities at $M_H=2M_{Z,W}$ are still noticeable, but the relativistic corrections make them far less pronounced. For $M_H$ away from that resonant area, relativistic corrections result in an increase of the branching fractions by factors ranging from about 2 to 10 and above, depending on the specific decay mode and the value of the Higgs mass. However, the decay probabilities remain fairly small, and further studies are needed to identify possible areas where these decays may be useful experimentally.

\section{Acknowledgements}
The authors would like to thank V.V. Braguta and A.K. Likhoded for useful discussions. This work was supported in part by Russian Foundation of Basic Research (under grant 07-02-00417), MK-110.2008.2, and Russian Science Support Foundation.

\clearpage\newpage

\clearpage\newpage

\end{document}